\pdfoutput=1
\documentclass[sigconf,screen]{acmart}

\copyrightyear{2026}
\acmYear{2026}
\setcopyright{cc}
\setcctype{by}
\acmConference[MM '26]{Proceedings of the 34th ACM International Conference on Multimedia}{November 10--14, 2026}{Rio de Janeiro, Brazil}
\acmBooktitle{Proceedings of the 34th ACM International Conference on Multimedia (MM '26), November 10--14, 2026, Rio de Janeiro, Brazil}
\acmDOI{10.1145/3767308.3835727}
\acmISBN{979-8-4007-2213-4/2026/11}
\usepackage{booktabs}
\usepackage{graphicx}
\usepackage{amsmath}

\usepackage{amssymb}
\usepackage{multirow}
\usepackage{balance}
\usepackage{xcolor}
\usepackage{algorithm}
\usepackage{algpseudocode}
\usepackage{tikz}
\usepackage{pgfplots}
\pgfplotsset{compat=1.18}
\usetikzlibrary{arrows.meta, positioning, shapes.geometric, fit, backgrounds, calc, patterns, decorations.pathreplacing}

\usepackage[subtle, mathdisplays=tight, lists=tight, indent=tight]{savetrees}

\usepackage{pifont}

\newcommand{\method}{NeuralLVC}

\newcommand{\eg}{\textit{e.g.}}
\newcommand{\etal}{\textit{et al.}}

\begin{document}

\title{\method: Neural Lossless Video Compression via Masked Diffusion with Temporal Conditioning}

\author{Tiberio Uricchio}
\affiliation{%
  \institution{University of Pisa}
  \city{Pisa}
  \country{Italy}
}
\email{tiberio.uricchio@unipi.it}

\author{Marco Bertini}
\affiliation{%
  \institution{Universit\`a degli Studi di Firenze}
  \city{Florence}
  \country{Italy}
}
\email{marco.bertini@unifi.it}

\renewcommand{\shortauthors}{Tiberio Uricchio and Marco Bertini}
%% No italics, no superscripts, not anonymous
%% Use footnote or author note to identify equal contribution, shared contribution, and/or contact author info

\begin{abstract}
% ===== BEGIN sections/abstract (flattened for camera-ready) =====
While neural lossless image compression has advanced significantly with learned entropy
models, lossless \emph{video} compression remains largely unexplored in the neural
setting. We present \method{}, a neural lossless video codec that combines masked
diffusion with an I/P-frame architecture for exploiting temporal redundancy.
Our I-frame model compresses individual frames using bijective linear
tokenization that guarantees exact pixel reconstruction.
The P-frame model compresses temporal differences between consecutive
frames, conditioned on the previous decoded frame via a lightweight
reference embedding that adds only 1.3\% trainable parameters. Group-wise decoding
enables controllable speed--compression trade-offs.
Our codec is lossless in the input domain: for video, it reconstructs YUV420
planes exactly; for image evaluation, RGB channels are reconstructed exactly.
Experiments on 9 Xiph CIF sequences and on 24 HD sequences at 720p and
1080p (Xiph, UVG, and the modern BVI-AOM benchmark) show that
\method{} outperforms H.264 and H.265 lossless by a significant margin 
at every resolution tested. We verify exact
reconstruction through end-to-end encode--decode testing with arithmetic coding.
These results suggest that masked diffusion with temporal conditioning is a
promising direction for neural lossless video compression.

% ===== END sections/abstract =====

\end{abstract}




\maketitle

% ===== BEGIN sections/introduction (flattened for camera-ready) =====
\section{Introduction}
\label{sec:introduction}

Lossless video compression underpins a range of professional
multimedia workflows where any deviation from the original signal
is unacceptable. In medical imaging, endoscopic and
surgical video recordings are archived alongside patient records
and reviewed under regulatory scrutiny~\cite{liu2017medical}:
compression artefacts can misrepresent the boundary of a lesion,
compromise a diagnostic AI system trained to detect tiny features,
or invalidate a medico-legal audit trail.
In broadcast and post-production, original camera files are passed
through multiple processing steps like grading, visual effects, and
finishing pipelines; each encode--decode cycle that introduces even
minor artefacts accumulates into visible quality loss that is
commercially unacceptable~\cite{smpte2042}.
In film mastering and digital preservation, studio archives and
national film institutions maintain lossless masters from which all
distribution formats are derived~\cite{rfc9043ffv1,kromer2017ffv1}:
the master must remain pixel-perfect through migrations and format
conversions. Across all of these settings, compression must reduce
storage and transmission cost while guaranteeing that every
reconstructed sample is identical to the original.

Traditional lossless video codecs like H.264 (Hi444PP profile) and H.265
(RExt profile) achieve compression through hand-crafted predictors and
entropy coding. These methods exploit spatial and temporal redundancy
using block-based motion estimation and transform coding. 

On the other hand, neural network approaches have dramatically advanced
\emph{lossy} video compression over the past decade. The DCVC
family~\cite{dcvc2021,dcvc-hem2022,dcvc-dc2023,dcvc-fm2024,
dcvc-rt2025} refined conditional coding with progressively richer
temporal context, with DCVC-HEM~\cite{dcvc-hem2022} being the
first learned codec to surpass H.266/VVC in rate--distortion
performance. Long-term temporal modelling~\cite{dcvclcg2024} and
generative latent coding~\cite{glc2024cvpr,glcvideo2025tcsvt} have
since pushed the frontier further. Yet the rate--distortion trade-off
introduced in these designs is fundamentally incompatible with the
exact-reconstruction requirement that medical and TV/film productions
demand.

On the \emph{lossless} side, neural compression has made
significant progress for still images. LC-FDNet~\cite{lcfdnet2022},
ArIB-BPS~\cite{aribbps2024}, CALLIC~\cite{callic2025},
FNLIC~\cite{fnlic2025}, and HPAC~\cite{hpac2025} now compete with
or surpass JPEG-XL on standard benchmarks. Large language models
have also emerged as effective entropy coders: LMCompress~\cite{lmcompress2025}
uses pretrained vision models with arithmetic coding and P2-LLM~\cite{p2llm2025}
frames next-pixel prediction as a language modelling task.
Despite this progress, these methods operate on individual still images. Lossless
compression of \emph{video}, where consecutive frames share
substantial temporal redundancy that could yield additional bitrate
savings, has not yet received enough attention in the neural setting.

In this paper, we present \method{}, a neural lossless video codec that
combines masked diffusion with an I/P-frame architecture to exploit temporal
redundancy in the lossless setting, based on two main components:

\textbf{Bidirectional Masked Diffusion.} Our entropy model uses
LLaDA~\cite{llada2025}, a bidirectional masked diffusion model that
conditions each prediction on \emph{all} unmasked positions. Combined
with HPAC group-wise parallel decoding~\cite{hpac2025}, this enables
efficient inference with controllable speed-quality trade-offs.

\textbf{I/P-Frame Architecture with Temporal Conditioning.} We introduce a
simple yet effective temporal framework: an I-frame model compresses individual
frames, while a P-frame model compresses temporal differences between consecutive
frames. The P-frame model is conditioned on the previous frame via a lightweight
reference embedding ($+$1.3\% parameters), enabling it to exploit temporal
redundancy. Both models use linear tokenization, a bijective mapping from pixel
values to tokens, ensuring exact reconstruction in the input domain.
This is a non-trivial constraint: a tokenisation that is too coarse
loses information, while one that is too fine inflates the vocabulary
and degrades the entropy model's probability estimates.

Our main contributions are:
\begin{itemize}
    \item To the best of our knowledge, we propose one of the \textbf{earliest
          temporally conditioned neural codecs for exact lossless video
          compression}, combining masked diffusion entropy modeling with
          an I/P-frame architecture.
    \item We demonstrate that \textbf{bijective linear tokenization} provides
          pixel-level lossless guarantees while enabling effective probability
          estimation through the masked diffusion framework.
    \item On 9 Xiph CIF sequences (YUV420), on three 720p Xiph
          sequences, on full-length 1080p, and on the UVG and BVI-AOM HD
          benchmarks (33 sequences in total), we
          \textbf{outperform H.264 and H.265 lossless} by a significant
          margin, and provide ablations isolating the contributions of
          temporal conditioning and reference embedding.
    \item We provide a \textbf{systematic comparison} against traditional codecs
          (H.264, H.265, VVC, FFV1, PNG), trivial diff-based baselines, and
          neural image methods, and verify losslessness through end-to-end
          encode-decode testing with arithmetic coding.
\end{itemize}

%\input{figures/pipeline}

% ===== END sections/introduction =====

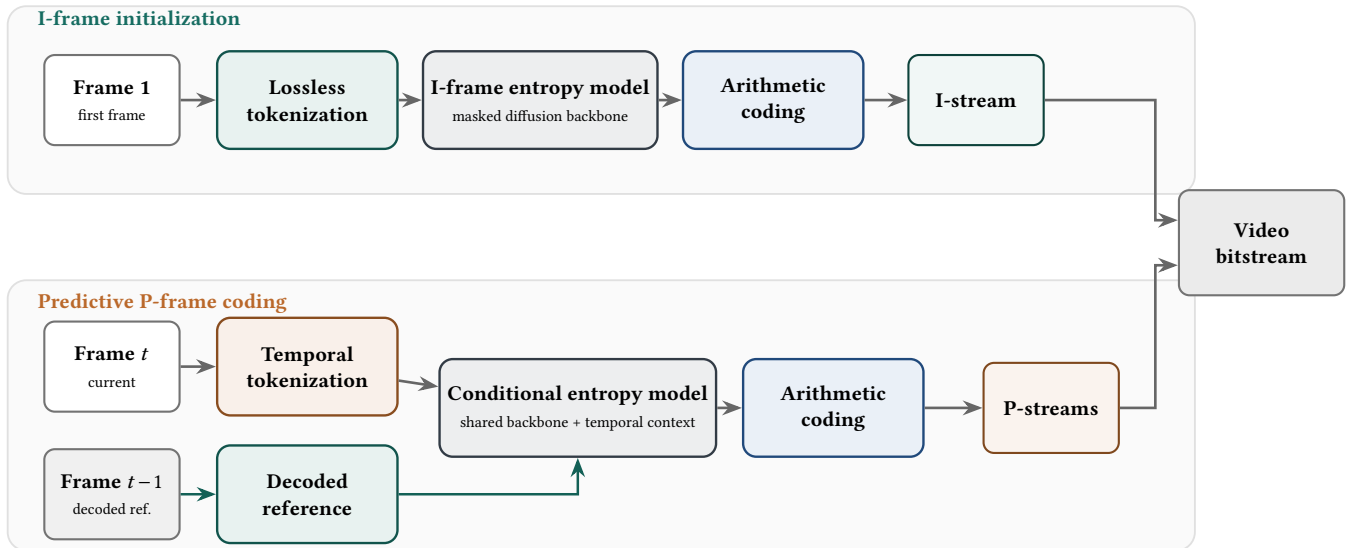
\begin{figure*}[t]
\centering
\definecolor{accentteal}{RGB}{24,120,108}
\definecolor{accentorange}{RGB}{196,110,44}
\definecolor{accentblue}{RGB}{46,104,176}
\definecolor{ink}{RGB}{72,84,98}
\resizebox{0.85\textwidth}{!}{%
\begin{tikzpicture}[
    >=Stealth,
    font=\small,
    panel/.style={
        draw=black!12,
        fill=black!2,
        rounded corners=7pt,
        line width=0.7pt
    },
    source/.style={
        draw=black!55,
        fill=white,
        rounded corners=4pt,
        minimum width=1.8cm,
        minimum height=1.2cm,
        align=center,
        line width=0.8pt,
        font=\small
    },
    stage/.style={
        draw=#1!70!black,
        fill=#1!10,
        rounded corners=5pt,
        minimum width=2.4cm,
        minimum height=1.3cm,
        align=center,
        line width=0.9pt,
        font=\small
    },
    badge/.style={
        draw=black!15,
        fill=white,
        rounded corners=4pt,
        inner xsep=6pt,
        inner ysep=3pt,
        font=\scriptsize
    },
    label/.style={font=\scriptsize\bfseries, text=black!70},
    arrow/.style={->, line width=1.0pt, draw=black!60},
    auxarrow/.style={->, line width=0.95pt, draw=accentteal!80!black},
]

% Panels
\node[panel, minimum width=15.8cm, minimum height=2.5cm, anchor=west] (topbg) at (-0.2,1.55) {};
\node[panel, minimum width=15.8cm, minimum height=3.6cm, anchor=west] (botbg) at (-0.2,-2.65) {};

% Titles
\node[anchor=west, font=\small\bfseries, text=accentteal!90!black] at (0.1,2.65) {I-frame initialization};
\node[anchor=west, font=\small\bfseries, text=accentorange!95!black] at (0.1,-1.15) {Predictive P-frame coding};

% Top lane
\node[source] (f1) at (1.2,1.55) {\textbf{Frame 1}\\{\scriptsize first frame}};
\node[stage=accentteal] (itok) at (3.8,1.55) {\textbf{Lossless}\\\textbf{tokenization}};
\node[stage=ink] (imodel) at (6.9,1.55) {\textbf{I-frame entropy model}\\{\scriptsize masked diffusion backbone}};
\node[stage=accentblue] (iac) at (10.0,1.55) {\textbf{Arithmetic}\\\textbf{coding}};
\node[source, fill=accentteal!6, draw=accentteal!55!black] (istream) at (12.7,1.55) {\textbf{I-stream}};

\draw[arrow] (f1) -- (itok);
\draw[arrow] (itok) -- (imodel);
\draw[arrow] (imodel) -- (iac);
\draw[arrow] (iac) -- (istream);

% Bottom lane sources
\node[source] (ft) at (1.2,-2.0) {\textbf{Frame $t$}\\{\scriptsize current}};
\node[source, fill=black!6] (fprev) at (1.2,-3.7) {\textbf{Frame $t\!-\!1$}\\{\scriptsize decoded ref.}};

% Bottom lane stages
\node[stage=accentorange] (resid) at (3.8,-2.0) {\textbf{Temporal}\\\textbf{tokenization}};
\node[stage=accentteal] (refctx) at (3.8,-3.7) {\textbf{Decoded}\\\textbf{reference}};
\node[stage=ink] (pmodel) at (7.4,-2.55) {\textbf{Conditional entropy model}\\{\scriptsize shared backbone + temporal context}};
\node[stage=accentblue] (pac) at (10.8,-2.55) {\textbf{Arithmetic}\\\textbf{coding}};
\node[source, fill=accentorange!8, draw=accentorange!65!black] (pstream) at (13.7,-2.55) {\textbf{P-streams}};

\draw[arrow] (ft) -- (resid);
\draw[auxarrow] (fprev) -- (refctx);
\draw[arrow] (resid) -- (pmodel);
\coordinate (refturn) at (pmodel.south |- refctx.east);
\draw[auxarrow] (refctx.east) -- (refturn) -- (pmodel.south);
\draw[arrow] (pmodel) -- (pac);
\draw[arrow] (pac) -- (pstream);

% Merge
\node[source, minimum width=2.2cm, minimum height=1.4cm, fill=black!8] (bitstream) at (16.5,-0.35) {\textbf{Video}\\\textbf{bitstream}};
\coordinate (iturn) at ([xshift=-0.3cm,yshift=0.3cm]bitstream.west);
\coordinate (pturn) at ([xshift=-0.3cm,yshift=-0.3cm]bitstream.west);
\draw[arrow] (istream.east) -- (iturn |- istream.east) -- (iturn) -- ([yshift=0.3cm]bitstream.west);
\draw[arrow] (pstream.east) -- (pturn |- pstream.east) -- (pturn) -- ([yshift=-0.3cm]bitstream.west);

\end{tikzpicture}%
}
\Description{Editorial overview of NeuralLVC with two horizontal lanes. The top lane compresses the first frame through lossless tokenization, an I-frame entropy model, arithmetic coding, and an I-stream. The bottom lane compresses later frames using the current frame plus the previous decoded frame, a conditional entropy model, arithmetic coding, and P-streams. Both lanes merge into one lossless video bitstream.}
\caption{High-level overview of \method{}. The first frame is coded independently, while later
frames are coded from the current frame together with the previous decoded frame. Both branches
use the same masked-diffusion entropy-modeling backbone and are finally compressed with arithmetic
coding into a single lossless bitstream. Exact token mappings and the group-wise decoding strategy
are described in the method section rather than embedded in the figure.}
\label{fig:pipeline}
\end{figure*}

% ===== BEGIN sections/related_work (flattened for camera-ready) =====
\section{Related Work}
\label{sec:related}

\subsection{Neural Lossless Image Compression}
Neural lossless image compression has advanced substantially in recent years.
PixelCNN~\cite{pixelcnn2016} introduced autoregressive image modeling by predicting each pixel conditioned on previously generated pixels. This formulation was later adapted for compression by using the predicted probability distributions for arithmetic coding. L3C~\cite{l3c2019} introduced hierarchical latent codes to enable partially parallel compression, trading some compression efficiency for higher speed.

More recent work has continued to improve the state-of-the-art in neural \emph{image} compression.
LC-FDNet~\cite{lcfdnet2022} proposed coarse-to-fine prediction with frequency decomposition, showing that multi-scale context can improve compression efficiency.
ArIB-BPS~\cite{aribbps2024} used bit-plane slicing with autoregressive models and reported strong results on standard benchmarks.
DLPR~\cite{dlpr2024} presented a lossy-plus-residual framework that first performs lossy compression and then losslessly encodes the residuals.
CALLIC~\cite{callic2025} combined masked convolutions with efficient cache-then-crop inference.
FNLIC~\cite{fnlic2025} introduced fitted neural compression with test-time adaptation to improve per-image compression.
HPAC~\cite{hpac2025} proposed hierarchical parallelism with group-wise autoregressive decoding, achieving state-of-the-art compression (2.52 bpsp on Kodak with fine-tuning) with only 677K parameters.
P2-LLM~\cite{p2llm2025} showed that large language models can act as entropy models for next-pixel prediction, while LLM-VP~\cite{llmvp2025} conditions LLMs with visual prompts for lossless coding.
Beyond pixel-domain compression, Li~\etal~\cite{li2025heicLVLM} proposed a token-level optimization strategy for compressing images intended for large vision-language models (LVLMs), jointly training a pre-editing network and an end-to-end codec using semantic token distortion and rank losses. Their method achieves more than 50\% bitrate savings relative to VVC when LVLMs are the ultimate receiver.
VoCo-LLaMA~\cite{vocollama2025} explored a complementary direction by distilling how LLMs process vision tokens into compact Vision Compression (\emph{VoCo}) tokens via attention masking, achieving a 576$\times$ compression ratio while retaining 83.7\% of task performance and reducing KV-cache storage by 99.8\%.

However, all of these methods focus on lossless or perceptual \emph{image} compression rather than lossless video compression.

Most relevant to our work, LMCompress~\cite{lmcompress2025} showed that pretrained vision models such as iGPT~\cite{igpt2020} can serve as strong entropy models for lossless compression. However, it processes frames independently and therefore does not exploit temporal redundancy. In addition, its cluster-based tokenization introduces quantization, which prevents pixel-level lossless reconstruction and yields only token-level losslessness.
Tsai~\cite{tsai2026llmcompression} revisited LLM-based lossless compression with arithmetic coding and showed that post-training quantization (HQQ at 3-bit) reduces the adjusted compression rate to 18\% on enwik9 without retraining. This result suggests that small, quantized language models are viable entropy coders, a direction that is orthogonal to, but compatible with, our temporally conditioned approach.

\subsection{Traditional Lossless Video Codecs}
The H.264/AVC~\cite{h264} and H.265/HEVC~\cite{h265} standards support lossless modes through their high-profile extensions. The H.264 Hi444PP profile enables lossless coding with 4:4:4 chroma sampling, while the H.265 RExt profile provides similar functionality with improved transform coding. Both rely on intra prediction, residual coding, and CABAC entropy coding. FFV1~\cite{ffv1} is a lossless codec designed primarily for archival use, favoring simplicity and speed over maximum compression. VVC/H.266 provides near-lossless compression at QP=0, but still introduces small quantization errors and is therefore not strictly lossless.
Although these traditional codecs achieve reasonable compression, they remain constrained by hand-crafted prediction models.

\subsection{Neural Lossy Video Compression}
Neural lossy video compression has progressed rapidly, driven by end-to-end rate-distortion optimization. DVC~\cite{lu2019dvc} pioneered an optical-flow residual-coding paradigm. The DCVC family~\cite{dcvc2021,dcvc-hem2022,dcvc-dc2023,dcvc-fm2024,dcvc-rt2025} extended this line of work through conditional coding with progressively richer temporal context.
DCVC-HEM~\cite{dcvc-hem2022} introduced hybrid spatial-temporal entropy modeling and was the first neural codec reported to surpass H.266 at the highest compression-ratio setting.
DCVC-DC~\cite{dcvc-dc2023} enriched both temporal and spatial context through hierarchical quality patterns and quadtree-based entropy coding.
DCVC-FM~\cite{dcvc-fm2024} introduced feature modulation to support a wider bitrate range, while DCVC-RT~\cite{dcvc-rt2025} achieved real-time 1080p coding at over 100 FPS by eliminating explicit motion modules.
Temporal context modeling was further extended in DCVC-LCG~\cite{dcvclcg2024}, which introduced a Long-term Context Gathering module to retrieve diverse temporal references beyond the immediately preceding frame, thereby mitigating error propagation in long prediction chains.
Related work on context mining also explored temporal context extraction across multi-scale propagated features~\cite{sheng2022tmm}.
Concurrently, GLC~\cite{glc2024cvpr} proposed compressing the VQ-VAE latent domain rather than pixel space to improve high-realism, high-fidelity reconstruction at ultra-low bitrates, with a video extension presented in~\cite{glcvideo2025tcsvt}.

\subsection{Masked Diffusion Models}
Discrete diffusion models~\cite{austin2021structured} extend diffusion processes to categorical data by progressively masking tokens and learning to reconstruct them. LLaDA~\cite{llada2025} applies this idea to language modeling, using \emph{bidirectional} attention instead of the causal (unidirectional) attention used in standard autoregressive models. As a result, each position can attend to all unmasked positions, yielding richer contextual information for prediction.

We adopt LLaDA as our entropy model because bidirectional attention is particularly well-suited for image compression: spatial dependencies in images are inherently non-causal, and attending to all spatial positions produces better probability estimates than left-to-right processing. Combined with HPAC's group-wise parallelism~\cite{hpac2025}, LLaDA enables efficient parallel prediction while maintaining high-quality probability distributions.

\subsection{Diffusion Models for Lossy Video Compression}
Diffusion models have also been explored extensively for \emph{lossy} and \emph{semantic} video compression, in which the decoder generates plausible frames from compact multimodal representations rather than reconstructing them exactly.

UQDM~\cite{uqdm2025} replaces Gaussian noise with uniform noise in diffusion models, enabling progressive coding from lossy to lossless with a single model.
While conceptually related, UQDM uses \emph{continuous} diffusion on latent representations, whereas our approach uses \emph{discrete} masked diffusion directly on pixel tokens, which naturally provides lossless guarantees.
CMVC~\cite{cmvc2025} encodes video as cross-modal text and keyframe representations using MLLMs and reconstructs via text-to-video or image-text-to-video generation, achieving competitive perceptual quality at ultra-low bitrates with no fidelity guarantees.
M3-CVC~\cite{m3cvc2025} similarly employs a dialogue-based large multimodal model (LMM) to extract hierarchical spatiotemporal descriptions, then uses conditional diffusion for keyframe and clip reconstruction, outperforming VVC at ultra-low bitrates on perceptual metrics.
DiffVC~\cite{Ma-2025} introduced a diffusion-based perceptual video compression framework that integrates a foundational diffusion model into the video conditional coding pipeline. The method uses temporal context from previously decoded frames and the current frame’s reconstructed latent representation to guide high-quality reconstruction, while Temporal Diffusion Information Reuse and quantization-parameter prompting improve inference efficiency and robustness across bitrates.
DiSCo~\cite{disco2025} factorizes video into a text description, a spatiotemporally degraded scaffold, and optional sketch or pose sequences, then reconstructs via a conditional video diffusion Transformer fine-tuned with in-context LoRA; it reports 5--10$\times$ better perceptual quality than traditional codecs and 2--3$\times$ better than prior semantic codecs at low bitrates.
CPSGD~\cite{cpsgd2026} further decomposes motion into camera-pose trajectories for the background and sparse foreground segmentation masks, guiding a fine-tuned Stable Video Diffusion-XL model to reconstruct video at approximately 0.003~BPP.
These generative approaches trade pixel-perfect fidelity for extreme compression and are therefore complementary to, rather than direct competitors with, our lossless objective.

\subsection{Neural Lossless Video Compression}
As the preceding subsections show, neural video compression research has focused primarily on \emph{lossy} compression, while neural lossless video compression remains largely unexplored.
LMCompress~\cite{lmcompress2025} briefly reports video results in its supplementary material, but it processes frames independently and does not model temporal redundancy.

To the best of our knowledge, our work is among the earliest to introduce a temporally conditioned neural codec for exact lossless video compression, with an I/P-frame architecture that explicitly models temporal redundancy.

% ===== END sections/related_work =====

% ===== BEGIN sections/method (flattened for camera-ready) =====
\section{Method}
\label{sec:method}

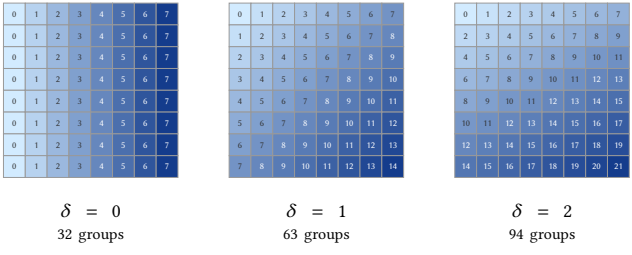
\begin{figure}[t]
\centering
\begin{tikzpicture}[
    cell/.style={minimum size=0.29cm, inner sep=0pt, outer sep=0pt,
                 draw=black!40, line width=0.15pt,
                 font=\sffamily\fontsize{3.5}{3.5}\selectfont},
    grouplabel/.style={font=\footnotesize, align=center, text width=2.5cm}
]

% Color gradient: light blue (group 0) to dark navy (max group).
\definecolor{grouplow}{RGB}{210,235,255}
\definecolor{grouphigh}{RGB}{20,60,140}

% Helper: pick white text for dark cells, dark text for light cells.
\newcommand{\cellnode}[3]{%
  % #1 = pct of grouplow (0=dark, 100=light), #2 = position, #3 = label
  \ifnum#1<45
    \node[cell, fill=grouplow!#1!grouphigh, text=white] at (#2) {#3};%
  \else
    \node[cell, fill=grouplow!#1!grouphigh, text=black!80] at (#2) {#3};%
  \fi
}

\newcommand{\cellsize}{0.29}

% --- Delta = 0 ---
% s(r,c) = c, groups 0..7 (8 groups)
\begin{scope}[xshift=0cm]
  \foreach \r in {0,...,7} {
    \foreach \c in {0,...,7} {
      \pgfmathtruncatemacro{\grp}{\c}
      \pgfmathtruncatemacro{\pct}{round(100 - 100*\grp/7)}
      \cellnode{\pct}{\c*\cellsize, -\r*\cellsize}{\grp}
    }
  }
  \node[grouplabel, anchor=north] at (3.5*\cellsize, -8.3*\cellsize)
    {$\delta=0$\\[0.5pt]{\scriptsize 32 groups}};
\end{scope}

% --- Delta = 1 ---
% s(r,c) = c + r, groups 0..14 (15 groups)
\begin{scope}[xshift=3.0cm]
  \foreach \r in {0,...,7} {
    \foreach \c in {0,...,7} {
      \pgfmathtruncatemacro{\grp}{\c + \r}
      \pgfmathtruncatemacro{\pct}{round(100 - 100*\grp/14)}
      \cellnode{\pct}{\c*\cellsize, -\r*\cellsize}{\grp}
    }
  }
  \node[grouplabel, anchor=north] at (3.5*\cellsize, -8.3*\cellsize)
    {$\delta=1$\\[0.5pt]{\scriptsize 63 groups}};
\end{scope}

% --- Delta = 2 ---
% s(r,c) = c + 2r, groups 0..21 (22 groups)
\begin{scope}[xshift=6.0cm]
  \foreach \r in {0,...,7} {
    \foreach \c in {0,...,7} {
      \pgfmathtruncatemacro{\grp}{\c + 2*\r}
      \pgfmathtruncatemacro{\pct}{round(100 - 100*\grp/21)}
      \cellnode{\pct}{\c*\cellsize, -\r*\cellsize}{\grp}
    }
  }
  \node[grouplabel, anchor=north] at (3.5*\cellsize, -8.3*\cellsize)
    {$\delta=2$\\[0.5pt]{\scriptsize 94 groups}};
\end{scope}

\end{tikzpicture}
\Description{Three 8x8 grids showing HPAC grouping patterns for delta values 0, 1, and 2, where each cell is colored by its group index and cells in the same group are predicted in parallel.}
\caption{Grouping patterns for different $\delta$ values on an 8$\times$8 grid (32$\times$32 in practice). Each color represents a group of positions predicted in parallel. $\delta=0$ yields column-wise groups; $\delta=1$ produces diagonal bands with more groups and better compression; $\delta=2$ creates steeper diagonals. The number in each cell indicates the group index.}
\label{fig:hpac_groups}\vspace{-12pt}
\end{figure}

We present \method{}, our framework for neural lossless video compression
(Figure~\ref{fig:pipeline}).
We describe the bijective tokenization that guarantees pixel-level losslessness,
the LLaDA masked diffusion entropy model with group-wise parallelism
(Figure~\ref{fig:hpac_groups}), and the I/P-frame architecture with
temporal conditioning.

\subsection{Problem Formulation}

Given a video $\mathcal{V} = \{f_1, f_2, \ldots, f_T\}$ consisting of $T$ frames,
our goal is to produce a compressed bitstream $\mathcal{B}$ that minimizes
$|\mathcal{B}|$ while ensuring \emph{exact} reconstruction: $\hat{\mathcal{V}} = \mathcal{V}$.

Following information theory, the optimal code length for a symbol $x$ given
its probability $p(x)$ is $-\log_2 p(x)$ bits. Better probability
estimation therefore leads to shorter codes and better compression.

Each frame is processed in YUV420 format: the Y (luminance) channel at full
resolution and U, V (chrominance) channels at half resolution, following the
standard used by traditional video codecs. Each channel is divided into
non-overlapping $32 \times 32$ patches and compressed independently.

\subsection{Lossless Tokenization}
\label{sec:tokenization}

A critical requirement for lossless compression is that the tokenization must be
\emph{bijective}: every distinct pixel value must map to a unique token, and the
inverse mapping must exactly recover the original pixel.
Cluster-based tokenizations (\eg, iGPT~\cite{igpt2020}) map pixels to the
nearest centroid in a learned palette. While this produces effective probability
estimates, the mapping is non-injective: many pixel values map to the same
cluster ID, making it impossible to recover exact pixel values from tokens alone.

We instead use bijective linear tokenization.
Both tokenization schemes (described below) produce tokens in the range
$[0, 510]$. The embedding table includes one additional entry for the
special mask token used by the diffusion process.

For I-frames, each pixel value $x \in [0, 255]$ is mapped as
$\text{Token}_I(x) = 2x$, producing 256 distinct even-valued tokens in
$\{0, 2, 4, \ldots, 510\}$.
The inverse is $x = \text{Token}_I / 2$, which is exact because only even
tokens are used.
For P-frames, consecutive frames typically differ by only a few pixel
values, so we encode the temporal difference:
$\text{Token}_P(x_t, x_{t-1}) = (x_t - x_{t-1}) + 255$,
mapping the difference range $[-255, +255]$ to tokens in $[0, 510]$.
The inverse is $x_t = \text{Token}_P - 255 + x_{t-1}$, which is exact
given the already-decoded previous pixel $x_{t-1}$.
Both tokenizations share the range $[0, 510]$, which allows warm-starting
the P-frame model from the I-frame weights during training.

\subsection{Masked Diffusion Entropy Model}
\label{sec:llada}

Our entropy model is inspired by LLaDA~\cite{llada2025}, a discrete masked
diffusion model originally proposed for language generation. We adapt its
core masked-prediction framework for image compression, with several
architectural modifications tailored to our setting.

Unlike autoregressive models that use causal (unidirectional) attention and
predict tokens left to right, our model uses \emph{bidirectional} attention
so that each position can attend to every unmasked position in the patch.
This is well suited to image data, where spatial dependencies run in all
directions rather than following a fixed raster order.

An important distinction is between the \emph{attention mechanism} and the
\emph{decoding protocol}. The attention is bidirectional: when predicting a
masked token, the model conditions on all currently unmasked positions
regardless of their spatial location, exploiting context from above, below,
left, and right simultaneously. The decoding protocol, however, is
sequential: groups of tokens are revealed one at a time
(Section~\ref{sec:delta}), so the encoder and decoder share the same
context at every step, required for correct arithmetic coding.
This combination gives us the best of both worlds: rich spatial context
within each prediction step, and a well defined sequential order for
lossless coding.

During training, a masking ratio $t$ is sampled uniformly from $[0,1]$,
and each of the 1024 tokens in a patch is independently replaced by the
special mask token with probability~$t$. The model receives this partially
masked sequence and outputs a probability distribution over the full
vocabulary at every position. The training loss is the cross-entropy between
the predicted distribution and the true token at each masked position,
weighted by $1/t$ so that all masking levels contribute equally in
expectation. This weighting ensures the model learns to predict well both
when very few tokens are visible (hard, high-$t$ regime) and when most of
the patch is already revealed (easy, low-$t$ regime).

Our backbone is a Transformer~\cite{vaswani2017attention} with
8~layers, hidden dimension 384, 6~attention heads, and 1024 positions
(one per pixel in a $32{\times}32$ patch), totaling 15.18M parameters.
Compared to the original LLaDA architecture, we use learned absolute
positional embeddings (suitable for our fixed-size $32{\times}32$ patches,
where relative position encoding offers no advantage over absolute),
standard LayerNorm, and GELU activations instead of RoPE, RMSNorm, and
SwiGLU. These choices simplify the implementation without affecting
compression quality for our small, fixed-resolution input.
Further details of how the trained model is used for sequential probability
estimation and arithmetic coding are given in Section~\ref{sec:delta}.

\subsection{Group-wise Parallelism}
\label{sec:delta}

Encoding a patch with the masked diffusion model requires obtaining a
probability distribution for each token, conditioned on the tokens that
are already known. A na\"ive approach would process the 1024 tokens one at
a time, requiring 1024 sequential forward passes. Since our Transformer
uses bidirectional (non-causal) attention, key--value caching is not
applicable: unmasking a new token changes the attention output at every
other position, invalidating any cached states. Groupwise parallel
decoding, introduced by HPAC~\cite{hpac2025}, reduces the number of
required passes by processing multiple tokens simultaneously.

Each pixel at position $(r,c)$ in a $P \times P$ patch is assigned to
group $s(r,c) = c + r \cdot \delta$, where $\delta$ is a user-chosen
parameter. With $\delta = 2$ and $P = 32$ the scheme produces 94~groups.
At encoding/decoding step $g$, the model observes all tokens from groups
$0,\ldots,g{-}1$ (already decoded) while positions in groups
$g,\ldots,G{-}1$ remain masked. The model then predicts the probability
distribution for every position in group~$g$, and arithmetic coding
encodes (or decodes) the corresponding tokens. Because encoder and decoder
condition on identical context at every step, they produce identical
probability distributions, which is the prerequisite for correct arithmetic
coding.

\subsection{I/P-Frame Architecture}
\label{sec:ipframe}

To exploit temporal redundancy, we adopt an I/P-frame architecture inspired
by traditional video codecs.

The first frame of a sequence is compressed independently by the I-frame
model, which processes each $32{\times}32$ patch through the masked
diffusion backbone described above. Every patch is tokenized with the
I-frame mapping ($\text{Token} = 2x$) and coded with groupwise parallel
decoding.

Subsequent frames are compressed by the P-frame model, which encodes the
temporal difference $\text{Token}_P = (x_t - x_{t-1}) + 255$
(Section~\ref{sec:tokenization}).
To give the model access to the spatial content of the previous frame, we
add a \emph{reference embedding} layer. This is a learned embedding
table that maps each possible reference token to a vector of the same
dimension as the token embedding (adding ${\sim}$197K parameters,
$+$1.3\%).
At each position, the reference token $\text{Token}_{ref} = 2\,x_{t-1}$
(the previous pixel, tokenized with the I-frame mapping) is looked up in
the reference embedding table, and the resulting vector is added to the
sum of the token embedding and the positional embedding:
\begin{equation}
h_i = \text{TokEmb}(\text{Token}_P^{(i)})
    + \text{PosEmb}(i)
    + \text{RefEmb}(\text{Token}_{ref}^{(i)}).
\end{equation}
The transformer layers then process $h_1,\ldots,h_{1024}$ exactly as in
the I-frame model. Because the reference embedding is the only
architectural difference, the P-frame model can be warm-started from the
trained I-frame weights: all shared parameters are copied, and only the
reference embedding is initialized randomly and learned during P-frame
training.

For color video, we follow the standard YUV420 format. The Y, U, and V
channels are read directly from native Y4M files and compressed
\emph{independently} with the same model. No retraining is needed for
chrominance channels, as all planes are 8-bit signals processed as
$32{\times}32$ patches.

% ===== END sections/method =====

% ===== BEGIN sections/experiments (flattened for camera-ready) =====
\section{Experiments}
\label{sec:experiments}

\subsection{Setup}

We evaluate on 9 Xiph.org~\cite{xiph} CIF test sequences
(352$\times$288, 90--300 frames each, 2300 frames total) using all frames
per sequence, and at high definition on four groups of HD material:
three 720p Xiph sequences (1280$\times$720, 100 frames), the four 1080p
Xiph sequences over their \emph{full} length (217--375 frames each), the
seven-sequence UVG benchmark~\cite{uvg2020} (1920$\times$1080, 100
frames), and the modern ten-sequence BVI-AOM database~\cite{bviaom2024}
(1920$\times$1088, 64 frames) --- 33 test sequences in total across four
resolutions.
All YUV420 data is read directly from native Y4M files to avoid
conversion artifacts.
To contextualize our spatial modeling, we also evaluate the I-frame model
on the Kodak image dataset (24 RGB images, 768$\times$512).

The I-frame model (15.18M parameters) is trained on
a 50K-image subset of ImageNet~\cite{imagenet} (20 epochs) and the P-frame model
(15.18M parameters) on Vimeo-90k~\cite{vimeo90k}
(${\sim}$58K 7-frame sequences, 15 epochs) with random crops, variable
frame gap (1--3 frames), and scale augmentation.
Both models are trained with AdamW (learning rate $10^{-4}$, weight decay
$0.01$) and batch size 64.
No test video is used during training.
Compression rate is
$\text{compressed bits} / \text{raw bits} \times 100\%$,
where raw bits equals $(|Y| + |U| + |V|) \times 8$ for YUV420 and
$H \times W \times 3$ sub-pixels for Kodak (measured in bits per
sub-pixel, bpsp).

We compare against traditional lossless codecs: PNG, FFV1~\cite{ffv1},
H.264 lossless\footnote{x264, \texttt{-qp 0 -preset veryslow}},
H.265 lossless\footnote{x265, \texttt{-x265-params lossless=1 -preset veryslow}},
H.265 intra-only\footnote{\texttt{keyint=1}, forcing all I-frames},
and VVC near-lossless\footnote{VVenC~\cite{vvenc},
\texttt{--qp 0 --qpa 0 --preset slower}; QP=0 is \emph{not} truly
lossless as it introduces small quantization errors: we measure
Y-PSNR of 57.7--58.9\,dB on 1080p content (B-frames coded at
QP${\approx}5$), versus zero pixel error for our codec}---as well as
a trivial temporal baseline that computes per-plane temporal differences
and compresses each residual frame as a 16-bit PNG (diff+PNG).
All experiments run on an NVIDIA GH200;
frames not divisible by 32 are padded with edge replication;
the GOP consists of one I-frame (the first frame) followed by all P-frames.

\subsection{Lossless Video Compression}

Table~\ref{tab:main_results} presents results on 9 CIF sequences
(YUV420, all frames, $\delta=2$).

\begin{table}[t]
\centering
\caption{Lossless video compression rates (\%, lower is better) on Xiph CIF,
YUV420, all frames. Best truly lossless result in \textbf{bold}.}
\label{tab:main_results}
\setlength{\tabcolsep}{2.5pt}
\renewcommand{\arraystretch}{1.05}
\footnotesize
\begin{tabular}{@{}l|r|cc|ccc|c|c@{}}
\toprule
\textbf{Video} & \textbf{Fr.} & \textbf{PNG} & \textbf{FFV1} & \textbf{H.264} & \textbf{H.265} & \textbf{H.265$_\text{I}$} & \textbf{VVC$^*$} & \textbf{Ours} \\
\midrule
akiyo          & 300 & 40.76 & 32.99 & 12.45 & 12.22 & 39.94 & 9.64 & \textbf{9.76} \\
bus            & 150 & 54.47 & 49.90 & 42.42 & 41.19 & 60.47 & 30.63 & \textbf{36.62} \\
container      & 300 & 48.69 & 43.24 & 29.03 & 28.91 & 50.60 & 22.40 & \textbf{23.75} \\
foreman        & 300 & 49.38 & 43.05 & 35.05 & 34.98 & 50.28 & 27.91 & \textbf{30.01} \\
coastguard     & 300 & 52.57 & 48.20 & 40.85 & 40.23 & 57.65 & 28.85 & \textbf{33.38} \\
football       & 260 & 49.14 & 42.87 & 41.79 & 42.97 & 54.25 & 30.52 & \textbf{31.93} \\
mobile         & 300 & 68.78 & 62.77 & 44.50 & 41.01 & 73.49 & 32.71 & \textbf{36.43} \\
stefan         &  90 & 55.40 & 50.02 & 43.94 & 43.51 & 61.59 & 32.71 & \textbf{35.55} \\
hall\_mon.     & 300 & 46.00 & 40.74 & 40.88 & 42.35 & 49.78 & 29.76 & \textbf{31.23} \\
\midrule
\textbf{Avg.}  & 2300 & 51.69 & 45.97 & 36.77 & 36.37 & 55.34 & 27.24 & \textbf{29.85} \\
\bottomrule
\end{tabular}

\vspace{2pt}
{\scriptsize\raggedright H.265$_\text{I}$ = intra-only.
$^*$VVC QP=0 is near-lossless (introduces quantization errors).\par}
\end{table}

\method{} achieves 29.85\% average compression rate on native YUV420 data,
outperforming H.265 lossless (36.37\%) by 17.9\% relative and H.264
lossless (36.77\%) by 18.8\% relative.
The improvement is consistent across all 9 sequences, from nearly-static
content (akiyo, 9.76\%) to high-motion scenes (mobile, 36.43\%).
VVC at QP=0 (27.24\%) achieves a lower rate on average, but it introduces
quantization errors and is therefore not truly lossless; on the
near-static sequence akiyo, our method nearly matches VVC
(9.76\% vs.\ 9.64\%).

These results show that a neural entropy model with temporal conditioning
can outperform all truly lossless traditional codecs on standard CIF
content. The gap relative to H.265 intra-only (55.34\%) further confirms
that our temporal framework contributes a large share of the overall gain.

\subsection[Scaling to High-Definition Video]{Scaling to High-Definition Video}

{We evaluate beyond CIF at 720p and 1080p, on both classic Xiph content
and modern HD benchmarks.

\textbf{720p.} On three 1280$\times$720 Xiph sequences (100 frames each;
Table~\ref{tab:720p}), NeuralLVC achieves 25.84\% on FourPeople (low-motion
videoconference), outperforming H.265 lossless (31.07\%) by 16.8\% relative
and approaching near-lossless VVC (25.09\%); on the high-motion park\_joy
(44.91\%) and stockholm (42.52\%) it again beats H.265.

\textbf{Full-length 1080p.} On the four 1080p Xiph sequences over their
\emph{full} length (217--375 frames; Table~\ref{tab:hd}), NeuralLVC is the
best truly-lossless codec on every sequence, averaging 29.46\% against
35.73\% (FFV1), 36.06\% (H.264), and 38.00\% (H.265): the relative margin
over H.265 (22.5\%) is \emph{larger} than at CIF (17.9\%), so the advantage
does not erode with resolution. The full-length runs also confirm the
absence of error accumulation, as expected from bit-exact reference
reconstruction.

\textbf{Modern HD benchmarks.} On UVG~\cite{uvg2020} (1920$\times$1080,
seven sequences, 100 frames; Table~\ref{tab:uvg}) NeuralLVC averages
30.34\% and is the best truly-lossless codec on all seven sequences; on the
modern BVI-AOM database~\cite{bviaom2024} (1920$\times$1088, ten sequences,
64 frames) it averages 24.1\% versus 28.4\% for H.265 lossless, again best
on every sequence.

Two observations follow. First, the rate tracks scene content rather than
resolution (29.85\% at CIF, {37.8\% over the three 720p
sequences, two of them high-motion,} and 29.46\% at
1080p). Second, at HD, H.264 overtakes H.265 among traditional lossless
codecs, since lossless coding bypasses the transform and quantization that
account for most of HEVC's lossy-mode gain~\cite{zhou2012hevc}, while our
margin persists (all H.264/H.265 streams decode-verified bit-exact).
Near-lossless VVC remains ahead (0.7--1.8 points on UVG and full-length
1080p Xiph, up to 2.5 on the 720p set), but that gap buys residual
distortion (measured Y-PSNR 57.7--58.9\,dB), not exact reconstruction.}

\begin{table}[t]
\centering
\caption{Lossless compression on 720p Xiph sequences (YUV420, 100 frames).
$^*$VVC QP=0 is near-lossless. Best truly lossless in \textbf{bold}.}
\label{tab:720p}
\small
\begin{tabular}{l|ccc|c}
\toprule
\textbf{Video} & \textbf{H.264} & \textbf{H.265} & \textbf{VVC$^*$} & \textbf{Ours} \\
\midrule
FourPeople    & 30.10 & 31.07 & 25.09 & \textbf{25.84} \\
park\_joy     & 49.19 & 48.53 & 41.97 & \textbf{44.91} \\
stockholm     & 46.92 & 46.93 & 38.74 & \textbf{42.52} \\
\bottomrule
\end{tabular}
\end{table}

\begin{table}[t]
\centering
\caption{Lossless compression on \emph{full-length} 1080p Xiph sequences
(YUV420, all frames). $^*$VVC QP=0 is near-lossless. Best truly lossless in
\textbf{bold}.}
\label{tab:hd}
\small
{
\begin{tabular}{l|r|cccc|c}
\toprule
\textbf{Video} & \textbf{Fr.} & \textbf{FFV1} & \textbf{H.264} & \textbf{H.265} & \textbf{VVC$^*$} & \textbf{Ours} \\
\midrule
blue\_sky    & 217 & 36.55 & 34.69 & 34.42 & 26.06 & \textbf{28.29} \\
pedestrian   & 375 & 31.48 & 32.06 & 33.63 & 24.47 & \textbf{25.98} \\
riverbed     & 250 & 37.93 & 40.63 & 46.65 & 32.11 & \textbf{33.87} \\
station2     & 313 & 36.95 & 36.85 & 37.31 & 28.12 & \textbf{29.70} \\
\midrule
\textbf{Avg.}&     & 35.73 & 36.06 & 38.00 & 27.69 & \textbf{29.46} \\
\bottomrule
\end{tabular}}
\end{table}

\begin{table}[t]
\centering
\caption{Per-sequence lossless rate (\% of raw, lower is better) on UVG
(1920$\times$1080, 100 frames). $^{*}$VVC QP=0 is near-lossless. Best
\emph{truly}-lossless in \textbf{bold}: ours wins all seven.}
\label{tab:uvg}
\footnotesize
{
\begin{tabular}{l ccccc}
\toprule
\textbf{UVG} & \textbf{FFV1} & \textbf{H.264} & \textbf{H.265} & \textbf{VVC$^{*}$} & \textbf{Ours} \\
\midrule
Beauty        & 43.63 & 45.10 & 45.79 & 40.19 & \textbf{40.31} \\
Bosphorus     & 29.59 & 29.24 & 29.30 & 22.55 & \textbf{23.16} \\
HoneyBee      & 37.35 & 37.04 & 37.63 & 31.57 & \textbf{31.79} \\
Jockey        & 33.63 & 34.94 & 35.38 & 28.67 & \textbf{30.44} \\
ReadySteadyGo & 35.17 & 32.75 & 32.64 & 26.27 & \textbf{27.50} \\
ShakeNDry     & 40.17 & 41.37 & 42.62 & 33.90 & \textbf{34.05} \\
YachtRide     & 30.82 & 32.01 & 32.93 & 24.55 & \textbf{25.16} \\
\midrule
\textbf{Avg.} & 35.77 & 36.06 & 36.61 & 29.67 & \textbf{30.34} \\
\bottomrule
\end{tabular}}
\end{table}

\subsection{Spatial Model vs.\ Image Codecs}
\label{sec:kodak}

To isolate the quality of our spatial entropy model from the temporal
component, we evaluate the I-frame model alone on the Kodak image
benchmark (Table~\ref{tab:kodak}).
Each RGB image is processed channel-wise (R, G, B independently).
We separate methods into pretrained models and those that perform
per-image fine-tuning during encoding.

\begin{table}[t]
\centering
\caption{Lossless image compression on Kodak (bpsp, lower is better).
All methods operate in the RGB domain.
}
\label{tab:kodak}
\small
\begin{tabular}{l|c|c}
\toprule
\textbf{Method} & \textbf{Venue} & \textbf{bpsp} \\
\midrule
\multicolumn{3}{l}{\emph{Traditional codecs:}} \\
\quad PNG             & --           & 4.35 \\
\quad JPEG-XL         & --           & 2.87 \\
\midrule
\multicolumn{3}{l}{\emph{Neural (pretrained, no fine-tuning):}} \\
\quad L3C             & CVPR'19      & 3.26 \\
\quad HPAC            & arXiv'25     & 2.73 \\
\quad ArIB-BPS        & CVPR'24      & 2.78 \\
\quad Ours (I-frame)  & --           & {4.32} \\
\midrule
\multicolumn{3}{l}{\emph{Neural (per-image fine-tuning):}} \\
\quad FNLIC  & CVPR'25      & 2.88 \\
\quad CALLIC & AAAI'25      & 2.54 \\
\quad HPAC-FT & arXiv'25   & 2.52 \\
\bottomrule
\end{tabular}
\end{table}

Among pretrained methods, our I-frame model (4.32~bpsp) achieves a
marginally lower average rate than PNG
(4.35) but falls behind image-specialized codecs such as HPAC~(2.73)
and ArIB-BPS~(2.78). This gap reflects deliberate design choices: we
use bijective linear tokenization for guaranteed pixel-level losslessness
and process $32{\times}32$ patches independently, whereas HPAC and
ArIB-BPS operate on full images with masked convolutions, giving them
access to long-range spatial context across the entire frame.
Our model is also smaller than ArIB-BPS (15.18M vs.\ 146.6M).

\textbf{Can stronger image codecs replace temporal modeling?}
A natural question is whether applying a state-of-the-art neural image
codec independently to each video frame would match or exceed our
temporal approach. To investigate this in the same domain, we evaluate
both our I-frame model and ArIB-BPS~\cite{aribbps2024} per-frame on
five CIF sequences in RGB ($\text{bpsp} = \text{bits per sub-pixel}$).
Our I-frame model achieves 4.71~bpsp while ArIB-BPS achieves
3.06~bpsp, a gap of 1.65~bpsp that reflects the stronger spatial
modeling of ArIB-BPS (full-image masked convolutions, 146.6M parameters)
compared to our patch-based Transformer (15.18M parameters).
Despite this weaker spatial model, our full I+P system with temporal
conditioning achieves 2.39~bpsp on YUV420 video, below
ArIB-BPS's 3.06~bpsp in RGB, even though ArIB-BPS has a much stronger
spatial model. While this cross-domain comparison is only suggestive
(YUV420 chroma subsampling may ease compression of U/V channels),
it indicates that temporal conditioning more than compensates for
a weaker spatial model, providing gains that per-frame image codecs
cannot access since they do not model temporal redundancy.

\subsection{Ablation Study}
\label{sec:ablations}

Table~\ref{tab:ablation} isolates the contribution of each component.

\begin{table}[t]
\centering
\caption{Ablation study (9 CIF, all frames, $\delta=2$).
All entries use the same native YUV420 data and evaluation protocol.}
\label{tab:ablation}
\small
\begin{tabular}{l|c}
\toprule
\textbf{Configuration} & \textbf{Rate \%} \\
\midrule
\multicolumn{2}{l}{\emph{Traditional references:}} \\
\quad diff + PNG (16-bit)           & 70.98 \\
\quad H.265 intra-only              & 55.34 \\
\quad H.265 inter                   & 36.37 \\
\midrule
\multicolumn{2}{l}{\emph{Ours --- component analysis:}} \\
\quad I-frame only                  & 47.07 \\
\quad I+P, diff only (no ref)       & 45.91 \\
\quad I+P, diff + ref conditioning  & \textbf{29.85} \\
\bottomrule
\end{tabular}
\end{table}

The diff+PNG baseline, which stores each frame's per-pixel temporal
difference as a 16-bit PNG, achieves only 70.98\%, far worse than both
our method and traditional video codecs.
This confirms that the temporal residual is not trivially compressible
and that the gains of our system stem from the learned entropy model.

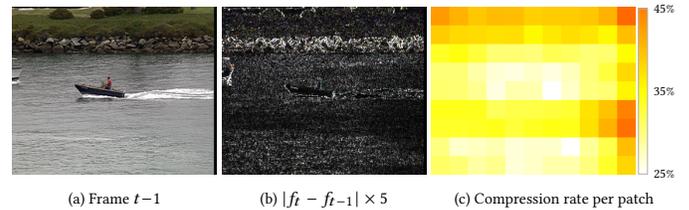
\begin{figure}[t]
\centering
\begin{tabular}{@{}c@{\hspace{2pt}}c@{\hspace{2pt}}c@{\hspace{1pt}}c@{}}
\includegraphics[width=0.32\columnwidth]{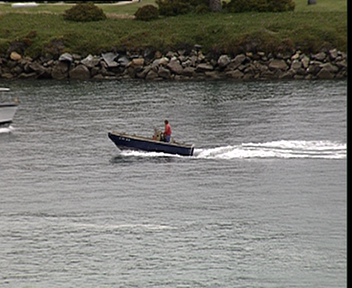} &
\includegraphics[width=0.32\columnwidth]{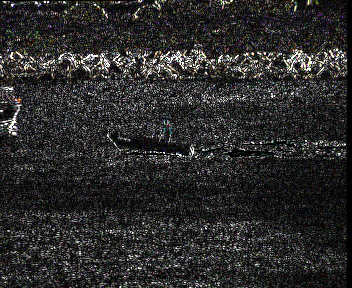} &
\includegraphics[width=0.32\columnwidth]{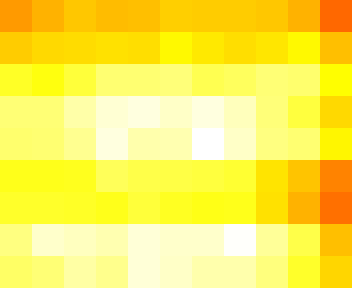} &
\begin{tikzpicture}[baseline=0cm]
  % Colorbar: white (low rate) at bottom → orange (high rate) at top
  \shade[bottom color=white, top color=yellow]
    (0,0) rectangle (0.12cm,1.1cm);
  \shade[bottom color=yellow, top color=orange]
    (0,1.1cm) rectangle (0.12cm,2.2cm);
  \draw[thin, gray!60] (0,0) rectangle (0.12cm,2.2cm);
  \node[font=\tiny, anchor=west, inner sep=1pt] at (0.16cm,2.2cm) {45\%};
  \node[font=\tiny, anchor=west, inner sep=1pt] at (0.16cm,1.1cm) {35\%};
  \node[font=\tiny, anchor=west, inner sep=1pt] at (0.16cm,0cm) {25\%};
\end{tikzpicture}
\\
{\scriptsize (a) Frame $t{-}1$} &
{\scriptsize (b) $|f_t - f_{t-1}| \times 5$} &
\multicolumn{2}{c}{\scriptsize (c) Compression rate per patch}
\end{tabular}
\Description{Three panels showing (a) a video frame, (b) the amplified temporal difference with the next frame, and (c) a heatmap of per-patch compression rate with colorbar where darker regions require more bits.}
\caption{Temporal redundancy and compression cost (coastguard, Y channel).
(a)~Reference frame. (b)~Temporal difference with the next frame
(amplified $5{\times}$): most change occurs along the moving boat.
(c)~Per-patch compression rate of the P-frame (dark = high rate, bright =
low rate): patches with large temporal differences require more bits,
while static regions compress to ${\sim}$28\%.}
\label{fig:visual_diff}
\end{figure}

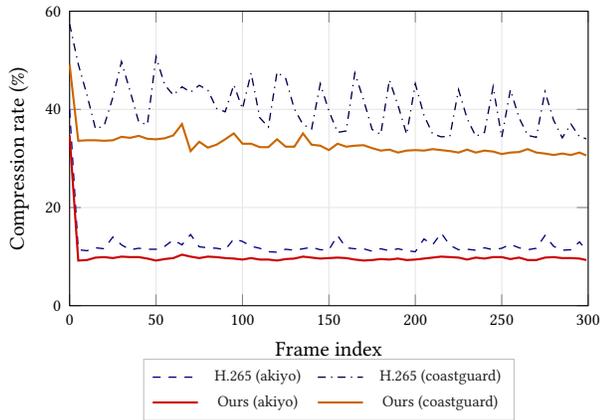
\begin{figure}[t]
\centering
\begin{tikzpicture}
\begin{axis}[
    width=0.9\columnwidth,
    height=5.5cm,
    xlabel={Frame index},
    ylabel={Compression rate (\%)},
    xmin=0, xmax=300,
    ymin=0, ymax=60,
    grid=major,
    grid style={gray!20},
    legend style={font=\scriptsize, draw=gray!50, fill=white,
                  fill opacity=0.9,
                  at={(0.5,-0.18)}, anchor=north,
                  legend columns=2, column sep=4pt},
    every axis label/.style={font=\small},
    tick label style={font=\scriptsize},
]

% H.265 akiyo — real per-frame (every 5 frames, 235B+63P+2I GOP)
\addplot[blue!50!black, line width=0.5pt, dashed]
    coordinates {
        (0,39.9) (5,11.4) (10,11.2) (15,11.8) (20,11.6) (25,14.0)
        (30,12.4) (35,11.4) (40,11.7) (45,11.5) (50,11.5) (55,12.1)
        (60,13.4) (65,12.4) (70,14.5) (75,12.0) (80,11.8) (85,11.7)
        (90,11.4) (95,13.6) (100,13.1) (105,12.1) (110,11.7) (115,11.0)
        (120,10.9) (125,11.5) (130,11.3) (135,11.6) (140,11.9) (145,11.4)
        (150,11.4) (155,14.3) (160,11.8) (165,11.6) (170,11.6) (175,11.1)
        (180,11.6) (185,11.2) (190,11.6) (195,11.2) (200,11.0) (205,13.6)
        (210,12.4) (215,14.8) (220,12.3) (225,11.4) (230,11.5) (235,11.3)
        (240,11.8) (245,11.4) (250,11.7) (255,12.5) (260,11.8) (265,11.4)
        (270,11.7) (275,14.3) (280,12.1) (285,11.3) (290,11.4) (295,13.0)
        (299,11.2)
    };
\addlegendentry{H.265 (akiyo)}

% H.265 coastguard — real per-frame
\addplot[blue!30!black, line width=0.5pt, dashdotted]
    coordinates {
        (0,57.4) (5,49.2) (10,43.0) (15,36.0) (20,36.7) (25,42.4)
        (30,49.7) (35,43.7) (40,37.4) (45,36.9) (50,50.7) (55,45.0)
        (60,42.9) (65,44.6) (70,43.5) (75,44.9) (80,43.9) (85,40.2)
        (90,39.5) (95,45.0) (100,40.1) (105,47.4) (110,38.3) (115,36.4)
        (120,47.6) (125,46.3) (130,40.1) (135,36.9) (140,36.0) (145,45.2)
        (150,39.7) (155,35.3) (160,35.6) (165,47.4) (170,42.0) (175,35.9)
        (180,34.9) (185,46.0) (190,40.8) (195,34.9) (200,45.2) (205,38.8)
        (210,35.0) (215,34.4) (220,34.5) (225,44.0) (230,38.1) (235,34.5)
        (240,35.3) (245,44.6) (250,34.3) (255,44.0) (260,38.3) (265,34.7)
        (270,34.3) (275,43.6) (280,37.8) (285,34.2) (290,37.0) (295,34.4)
        (299,34.0)
    };
\addlegendentry{H.265 (coastguard)}

% Ours akiyo — real per-frame (every 5 frames)
\addplot[red!80!black, line width=0.8pt]
    coordinates {
        (0,34.7) (5,9.2) (10,9.3) (15,9.8) (20,9.9) (25,9.7)
        (30,10.0) (35,9.9) (40,9.9) (45,9.6) (50,9.2) (55,9.5)
        (60,9.7) (65,10.4) (70,10.0) (75,9.7) (80,10.0) (85,9.9)
        (90,9.7) (95,9.6) (100,9.4) (105,9.7) (110,9.4) (115,9.4)
        (120,9.2) (125,9.5) (130,9.6) (135,10.0) (140,9.8) (145,9.6)
        (150,9.7) (155,9.8) (160,9.7) (165,9.4) (170,9.2) (175,9.3)
        (180,9.5) (185,9.4) (190,9.6) (195,9.3) (200,9.4) (205,9.6)
        (210,9.8) (215,10.0) (220,9.9) (225,9.8) (230,9.4) (235,9.8)
        (240,9.6) (245,9.9) (250,9.9) (255,9.5) (260,9.8) (265,9.3)
        (270,9.3) (275,9.8) (280,9.9) (285,9.7) (290,9.7) (295,9.6)
        (299,9.3)
    };
\addlegendentry{Ours (akiyo)}

% Ours coastguard — real per-frame
\addplot[orange!80!black, line width=0.8pt]
    coordinates {
        (0,49.1) (5,33.6) (10,33.7) (15,33.7) (20,33.6) (25,33.7)
        (30,34.4) (35,34.2) (40,34.6) (45,34.0) (50,33.9) (55,34.1)
        (60,34.7) (65,37.0) (70,31.5) (75,33.4) (80,32.2) (85,32.8)
        (90,33.9) (95,35.1) (100,33.0) (105,33.0) (110,32.3) (115,32.3)
        (120,33.9) (125,32.4) (130,32.4) (135,35.1) (140,32.8) (145,32.6)
        (150,31.7) (155,33.0) (160,32.4) (165,32.6) (170,32.7) (175,32.1)
        (180,31.6) (185,31.8) (190,31.2) (195,31.6) (200,31.7) (205,31.6)
        (210,31.9) (215,31.7) (220,31.5) (225,31.2) (230,31.8) (235,31.2)
        (240,31.6) (245,31.4) (250,30.9) (255,31.2) (260,31.3) (265,31.9)
        (270,31.2) (275,31.0) (280,30.7) (285,31.0) (290,30.7) (295,31.2)
        (299,30.6)
    };
\addlegendentry{Ours (coastguard)}

\end{axis}
\end{tikzpicture}
\Description{Line plot showing per-frame compression rate over 300 frames for our method and H.265 on akiyo and coastguard. Our method produces nearly constant P-frame rates while H.265 oscillates due to B-frame GOP structure.}
\caption{Per-frame compression rate on two CIF sequences with different
motion levels (YUV420, all 300 frames, sampled every 5).
Our method (solid) produces stable P-frame rates that consistently
outperform H.265 lossless (dashed). H.265 exhibits large per-frame
fluctuations due to its B-frame GOP structure (235~B, 63~P, 2~I frames),
while our codec maintains near-constant rates without error accumulation.}
\label{fig:per_frame_rate}\vspace{-12pt}
\end{figure}

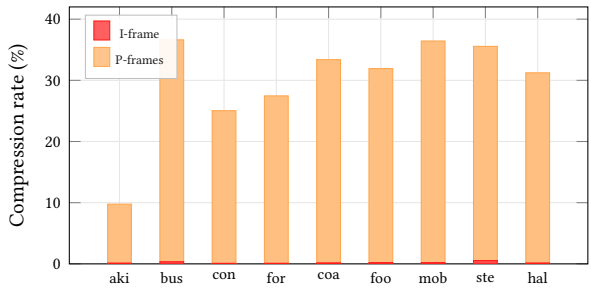
\begin{figure}[t]
\centering
\begin{tikzpicture}
\begin{axis}[
    width=0.9\columnwidth,
    height=5cm,
    ybar stacked,
    bar width=9pt,
    ylabel={Compression rate (\%)},
    symbolic x coords={aki,bus,con,for,coa,foo,mob,ste,hal},
    xtick=data,
    x tick label style={font=\scriptsize},
    ymin=0, ymax=42,
    grid=major,
    grid style={gray!20},
    legend pos=north west,
    legend style={font=\tiny, draw=gray!50, fill=white, fill opacity=0.9},
    every axis label/.style={font=\small},
    tick label style={font=\scriptsize},
    enlarge x limits=0.12,
]

% I-frame contribution = I-rate / T
% total = I_contribution + P_contribution
% I_contribution = I_rate * 1/T, P_contribution = P_rate * (T-1)/T
% From eval: akiyo total=9.76, I=34.70, T=300 → I_part=34.70/300=0.12
%   container: I=44.12 → 0.15, P_part=23.75-0.15=23.60
%   foreman:   I=40.29 → 0.13, P_part=30.01-0.13=29.88
\addplot[fill=red!70, draw=red!90] coordinates {
    (aki, 0.12) (bus, 0.34) (con, 0.15) (for, 0.13)
    (coa, 0.16) (foo, 0.20) (mob, 0.20) (ste, 0.54) (hal, 0.14)
};
\addlegendentry{I-frame}

% P-frame contribution = total - I_part
\addplot[fill=orange!50, draw=orange!70] coordinates {
    (aki, 9.64) (bus, 36.28) (con, 23.60) (for, 29.88)
    (coa, 33.22) (foo, 31.73) (mob, 36.23) (ste, 35.01) (hal, 31.09)
};
\addlegendentry{P-frames}

\end{axis}
\end{tikzpicture}
\Description{Stacked bar chart showing I-frame and P-frame contributions to total compression rate. I-frame contributes less than 1\% on all videos.}
\caption{Rate composition per video. The I-frame cost (dark) is amortized
over $T$~frames and contributes less than 1\% to the total. Compression
is dominated by P-frame performance.}
\label{fig:rate_composition}
\end{figure}

Temporal conditioning is the dominant factor: adding P-frame diff tokenization with
reference embedding conditioning reduces the rate from 47.07\% (I-frame
only) to 29.85\% (I+P conditioned), a 37\% relative improvement.
Without reference conditioning, the diff-only variant achieves 45.91\%,
only marginally better than I-frame only (47.07\%), confirming that the
reference embedding is essential for exploiting temporal redundancy.
Figure~\ref{fig:visual_diff} illustrates why temporal conditioning is
effective: consecutive frames differ primarily along moving edges, so the
temporal residual has much lower entropy than the raw frame.
Figure~\ref{fig:per_frame_rate} shows that our P-frame rates stabilize
quickly and remain nearly constant over 300 frames (std $<$1.6\%) without
error accumulation. In contrast, H.265 exhibits large periodic
fluctuations (std up to 5.0\%) caused by its B-frame GOP structure:
P-frames require more bits than B-frames (which exploit bidirectional
references), producing a regular oscillating pattern.
Because reconstruction is bit-exact, the P-frame reference never
drifts: over the 217--375-frame full-length runs of
Table~\ref{tab:hd} there is no error accumulation, and per-frame rate
tracks scene content rather than frame index, so per-frame cost does
not grow with sequence length, unlike lossy GOP structures.
Figure~\ref{fig:rate_composition} confirms that the I-frame cost is
amortized to less than 1\% of the total rate on long sequences.

\subsection{Speed Analysis}

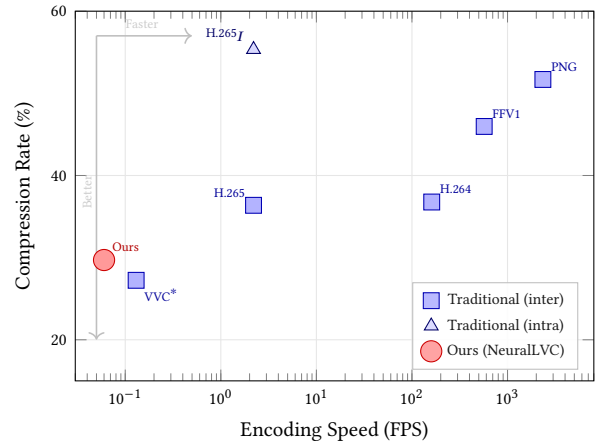
\begin{figure}[t]
\centering
\begin{tikzpicture}
\begin{axis}[
    width=0.9\columnwidth,
    height=6.5cm,
    xlabel={Encoding Speed (FPS)},
    ylabel={Compression Rate (\%)},
    xmode=log,
    xmin=0.03, xmax=8000,
    ymin=15, ymax=60,
    grid=major,
    grid style={gray!20},
    legend pos=south east,
    legend style={font=\scriptsize, draw=gray!50, fill=white, fill opacity=0.9},
    every axis label/.style={font=\small},
    tick label style={font=\scriptsize},
]

% Traditional codecs inter (squares) -- MEASURED on GH200, native Y4M
\addplot[only marks, mark=square*, mark size=3pt, blue!70!black, fill=blue!30]
    coordinates {
        (2344, 51.69)
        (570, 45.97)
        (161, 36.77)
        (2.2, 36.37)
        (0.13, 27.24)
    };
\addlegendentry{Traditional (inter)}

% Traditional intra (triangles)
\addplot[only marks, mark=triangle*, mark size=3pt, blue!40!black, fill=blue!15]
    coordinates {
        (2.2, 55.34)
    };
\addlegendentry{Traditional (intra)}

% (FNLIC removed: operates in RGB domain, not directly comparable to YUV420)

% Ours (circles, filled red)
\addplot[only marks, mark=*, mark size=4pt, red!80!black, fill=red!40]
    coordinates {
        (0.06, 29.85)
    };
\addlegendentry{Ours (\method{})}

% Labels for traditional
\node[font=\tiny, anchor=south west, text=blue!60!black] at (axis cs:2344,51.69) {PNG};
\node[font=\tiny, anchor=south west, text=blue!60!black] at (axis cs:570,45.97) {FFV1};
\node[font=\tiny, anchor=south west, text=blue!60!black] at (axis cs:161,36.77) {H.264};
\node[font=\tiny, anchor=south east, text=blue!60!black] at (axis cs:2.2,36.37) {H.265};
\node[font=\tiny, anchor=north west, text=blue!60!black] at (axis cs:0.13,27.24) {VVC$^*$};
\node[font=\tiny, anchor=south east, text=blue!40!black] at (axis cs:2.2,55.34) {H.265$_I$};

% Label for ours
\node[font=\tiny, anchor=south west, text=red!70!black] at (axis cs:0.06,29.85) {Ours};

% "Better" arrow
\draw[->, line width=0.6pt, gray!50] (axis cs:0.05,57) -- (axis cs:0.05,20);
\node[font=\tiny, text=gray!40, rotate=90] at (axis cs:0.04,38) {Better};
\draw[->, line width=0.6pt, gray!50] (axis cs:0.05,57) -- (axis cs:0.5,57);
\node[font=\tiny, text=gray!40] at (axis cs:0.15,58.5) {Faster};

\end{axis}
\end{tikzpicture}
\Description{Scatter plot of compression rate vs encoding speed for traditional codecs and NeuralLVC. NeuralLVC achieves 29.85\% at 0.06 FPS, between H.265 (36.37\% at 2.2 FPS) and VVC (27.24\% at 0.13 FPS).}
\caption{Compression rate vs.\ encoding speed on Xiph CIF (YUV420, native data).
All speeds measured on NVIDIA GH200.
$^*$VVC QP=0 is near-lossless.}
\label{fig:rate_speed}
\vspace{-12pt}
\end{figure}

Figure~\ref{fig:rate_speed} shows the compression speed trade-off, with
all speeds measured on the same hardware (NVIDIA GH200).
Our method operates at approximately 0.06~FPS end-to-end on CIF
with $\delta=2$, comparable in order of magnitude to VVC with the
\texttt{slower} preset (0.13~FPS).
H.265 lossless with the \texttt{veryslow} preset achieves only 2.2~FPS.
We chose \texttt{veryslow} / \texttt{slower} presets for the traditional
codecs to give them every advantage in compression: even in lossless mode,
slower presets find better intra/inter predictions, producing smaller
residuals and thus smaller bitstreams (up to ${\sim}$3.5\% difference
between \texttt{ultrafast} and \texttt{veryslow} for H.265 lossless).

The number of sequential forward passes equals the number of groups (94 for $\delta=2$, 63 for $\delta=1$), each requiring full
self-attention over all 1024 positions.
This is roughly $11\times$ fewer passes than token by token sequential
decoding (1024 passes). Because our model uses bidirectional attention,
standard key--value caching---which relies on each token attending only
to earlier positions---is not applicable: revealing a new token changes
the attention output at every other position.
Groupwise parallel decoding is therefore the natural acceleration
mechanism for masked diffusion entropy models.

% We chose masked diffusion over autoregressive (causal) modeling because
% bidirectional attention conditions each prediction on spatial context
% from \emph{all} directions within the patch, not only the raster-order
% ``left'' context available to causal models.
% For 2D image data this richer conditioning yields more accurate
% probability estimates, which translates directly into better
% compression.

Decoding speed is symmetric with encoding: the decoder performs the
same sequence of forward passes to reconstruct the probability
distributions, then applies arithmetic decoding. On CIF, the end-to-end codec including arithmetic
coding operates at ${\sim}$0.06~FPS with a GPU memory footprint of
${\sim}$4.5\,GB.
These results are encouraging for an initial system; improving speed
through architectural optimizations, speculative decoding, or
distillation is a promising direction for future work.

\textbf{Computational complexity.} One forward pass over a 1024-token
$32{\times}32$ patch costs 42.3~GFLOPs (21.1~GMACs); the 15.18\,M parameters are
shared and amortized across all sequences. A 1080p frame contains
${\sim}2{,}970$ YUV420 patches, so causal groupwise decoding ($\delta{=}2$, 94
groups) issues ${\sim}2.8{\times}10^{5}$ forward passes (${\sim}11.8$~PFLOPs) per
frame. Encoding time splits into 78\% probability estimation and 22\% arithmetic
coding, with measured wall-clock of 14.3\,s per CIF frame and 205--400\,s per
1080p frame. The cost is thus dominated by the number of sequential group passes
rather than raw FLOPs: compilation and speculative shortcuts give little
per-frame speedup, and while half-precision inference reduces the per-pass cost
(${\sim}3\times$, bit-exact), the sequential group structure remains the
dominant term; GPU memory is not a bottleneck (4{K} fits on a single GPU).

\subsection{Lossless Verification}
\label{sec:verification}

We implement a complete codec with arithmetic coding and verify
pixel-perfect reconstruction by encoding and decoding full video
sequences: the comparison between original and reconstructed frames
produces zero pixel error on all tested sequences.

This property distinguishes our method from VVC at QP=0, which is
often described as ``near-lossless'' but does introduce quantization
errors. Setting the base QP to~0 in VVenC does not produce truly
lossless output: the standard hierarchical GOP structure applies
QP offsets to inter-predicted frames, resulting in average B-frame
QP of ${\sim}$5--6 on our sequences and Y-channel PSNR of 57--61\,dB
(high, but not infinite). These errors may be negligible for display,
but they preclude domains that mandate bit-exact reconstruction,
such as long-term archival and preservation masters~\cite{ffv1} and
scientific or remote-sensing imagery, where lossless coding is the
operational standard~\cite{ccsds123review2021}.

% ===== END sections/experiments =====

% ===== BEGIN sections/conclusion (flattened for camera-ready) =====
\section{Conclusion}
\label{sec:conclusion}

We presented \method{}, a neural lossless video codec based on masked
diffusion models with an I/P-frame architecture that exploits temporal
redundancy. It uses bijective linear tokenization to guarantee
pixel-level lossless reconstruction in the input domain, and temporal
conditioning via a lightweight reference embedding.

On 9 Xiph CIF sequences (all frames, YUV420), \method{} achieves 29.85\%
average compression rate, outperforming H.265 lossless (36.37\%) by 17.9\%
relative and H.264 lossless (36.77\%) by 18.8\% relative
(Table~\ref{tab:main_results}).
The advantage carries to high definition without retraining:
\method{} is the best truly-lossless codec on every HD sequence tested
(Xiph, UVG, BVI-AOM), with the margin over H.265 growing to 22.5\% on
full-length 1080p Xiph.
The ablation (Table~\ref{tab:ablation}) confirms temporal conditioning
as the dominant factor in our gains.

Our method is slower than traditional codecs, limiting it to offline
archival---one of the most common scenarios for lossless video
compression. The I/P-frame structure assumes sequential frame access
without scene-change detection; periodic I-frame insertion mitigates but
does not fully address arbitrary scene cuts. Each $32{\times}32$ patch
is processed independently, without cross-patch spatial context or
cross-channel dependencies in YUV420.
\vspace{-1em}

% ===== END sections/conclusion =====

\begin{acks}
Work partially supported by the Italian Ministry of Education and Research (MUR) in the framework of the FoReLab project (Departments of Excellence). Part of the computational resources were provided by the Computing@UniPi, infrastructure of the University of Pisa (Green Data Center). Research supported by the NVIDIA Academic Grant Program using an NVIDIA DGX Spark system, CUDA, and NVIDIA NGC PyTorch containers.
\end{acks}

\balance
\bibliographystyle{ACM-Reference-Format}
\bibliography{references}

\end{document}